\documentclass[11pt,twoside]{article}
\usepackage{amssymb}
\usepackage{amsthm}
\usepackage{times,cite,bm}
\newcommand{\be}{\begin{eqnarray}}
\newcommand{\ee}{\end{eqnarray}}
\newcommand{\bez}{\begin{eqnarray*}}
\newcommand{\eez}{\end{eqnarray*}}
\newcommand{\pa}{\partial}
\newcommand{\A}{\mathbb{A}}
\newcommand{\cA}{\mathcal{A}}
\newcommand{\QSym}{\mathrm{QSym}}
\newcommand{\Sym}{\mathrm{Sym}}
\newcommand{\bu}{\bullet}

\theoremstyle{plain}
\newtheorem{theorem}{Theorem}[section]
\newtheorem{lemma}[theorem]{Lemma}
\newtheorem{proposition}[theorem]{Proposition}

\theoremstyle{definition}
\newtheorem{definition}[theorem]{Definition}
\newtheorem{remark}[theorem]{Remark}

\title{\bf Quasi-symmetric functions and the KP hierarchy
\thanks{\copyright 2009 by A. Dimakis and F. M\"uller-Hoissen} }

\author{Aristophanes Dimakis \\
 Department of Financial and Management Engineering, \\
 University of the Aegean, 31 Fostini Str., GR-82100 Chios, Greece \\
 dimakis@aegean.gr
          \and
 Folkert M\"uller-Hoissen \\ Max-Planck-Institute for Dynamics and Self-Organization \\
 Bunsenstrasse 10, D-37073 G\"ottingen, Germany \\
 folkert.mueller-hoissen@ds.mpg.de }

\date{ }

\begin{document}

\maketitle

\begin{abstract}
Quasi-symmetric functions show up in an approach to solve the Kadomtsev-Petviashvili (KP) hierarchy. 
This moreover features a new nonassociative product of quasi-symmetric functions that satisfies simple 
relations with the ordinary product and the outer coproduct. In particular, 
supplied with this new product and the outer coproduct, the algebra of quasi-symmetric functions 
becomes an infinitesimal bialgebra. Using these results we derive a sequence of identities 
in the algebra of quasi-symmetric functions that are in formal correspondence with the equations 
of the KP hierarchy. 
\end{abstract}

\section{Introduction}
\label{sec:intro}
Quasi-symmetric functions \cite{Gess84,Reut93,Malv+Reut95,Ehre96,GKLLRT95,Stan99v2,Haze03,Haze05} 
in a set of commuting variables extend the ring of symmetric functions \cite{Macd95} and 
show up in various branches of mathematics, most notably in combinatorics.
The theory of symmetric functions has many applications in mathematics and physics. 
In particular, Schur polynomials play an important role in the $\tau$-function formulation of the 
famous Kadomtsev-Petviashvili (KP) hierarchy (see e.g. \cite{MJD00}) of completely integrable 
nonlinear partial differential equations. 
In this work we demonstrate the appearance of \emph{quasi}-symmetric functions
for the potential form of the KP hierarchy, and more generally 
for its noncommutative version (see e.g. \cite{Kupe00}), where the 
dependent variable has values in a noncommutative associative algebra (typically an algebra 
of matrices of functions). Moreover, we show that the noncommutative KP hierarchy has an 
algebraic counterpart in the algebra of quasi-symmetric functions. 
These results involve a nonassociative product of quasisymmetric functions that satisfies 
nice relations with the ordinary product and the outer coproduct. Here we meet a weak form 
of nonassociativity characterized as follows \cite{DMH06nahier,DMH08isl}. 

\begin{definition}
Let $\A$ be a nonassociative ring (or algebra over a commutative ring) with product $\bu$. 
$\A$ is called \emph{weakly nonassociative} if\footnote{This 
appeared as rule T31R=T13L in a classification of weakenings of the associativity law \cite{Kunen96}. 
See also \cite{Pallo09} for some interesting relations. }
\be
    (a, b \bu c, d) = 0 \qquad \quad \forall a,b,c,d \in \A \, ,  \label{weak_nonassoc}
\ee
where $(a,b,c) = (a \bu b) \bu c - a \bu (b \bu c)$ is the associator. 
\end{definition}

Section~\ref{sec:KP->qs} demonstrates the appearance of quasi-symmetric functions, and the new product $\bu$ 
of the latter, in an approach to solve the KP hierarchy. 
Section~\ref{sec:qs} recalls some basic facts on quasi-symmetric functions needed in the sequel. 
In section~\ref{sec:new_products} we study the new product in more detail. 
In particular, we find that the algebra $\QSym$ of quasi-symmetric functions is generated by the 
identity element via $\bu$. Section~\ref{sec:relations} establishes relations between the new product 
and the outer coproduct \cite{Malv+Reut95}, which we denote by $\Delta$. 
In particular, $(\QSym,\bu, \Delta)$ turns out to be an \emph{infinitesimal bialgebra} 
\cite{Joni+Rota79,Agui00}. 
Section~\ref{sec:antipode} reveals a simple relation between the product $\bu$ and the antipode of the Hopf algebra 
of quasi-symmetric functions. Section~\ref{sec:KP-ids} derives a sequence of ``KP identities" in $\QSym$. 

The analysis of the KP equation (and more generally the KP hierarchy) in section~\ref{sec:KP->qs} actually 
exhibits a generalization of quasi-symmetric functions to \emph{quasi-supersymmetric} 
functions, and section~\ref{sec:qss} briefly treats this extension. The algebra of quasi-supersymmetric functions 
extends the algebra of supersymmetric functions \cite{MNR81,Bala+Bars81,Scheu84,Stem85,Macd95}, and 
an appearance of the latter in the context of the KP hierarchy has been 
noted in \cite{Nimm88,Nimm89}. Section~\ref{sec:final} contains some concluding remarks.

\section{From the KP equation to quasi-symmetric functions}
\label{sec:KP->qs}
The noncommutative KP equation is the partial differential equation
\bez
    4 \, \phi_{t_1 t_3} - 3 \, \phi_{t_2t_2} - \phi_{t_1t_1t_1t_1} 
  = 6 \, (\phi_{t_1} \phi_{t_1})_{t_1} - 6 \, [\phi_{t_1},\phi_{t_2}] \, ,  \label{KPeq}
\eez
where $\phi$ has values in a (typically non-commutative) associative algebra $\cA$, supplied with 
a structure to define differentiability with respect to independent variables $t_n$, $n=1,2,3$. 
$\phi_{t_n}$ denotes the partial derivative of $\phi$ with respect to $t_n$. 
Inserting the (formal) power series ansatz \cite{Okhu+Wada83,Kupe00,Pani01,DMH05KPalgebra}
\bez
    \phi = \sum_{n \geq 1} \epsilon^n \phi^{(n)} \, , 
\eez
and reading off coefficients of powers of the parameter $\epsilon$, we obtain the equations
\be
    4 \, \phi^{(n)}_{t_1 t_3} - 3 \, \phi^{(n)}_{t_2t_2} - \phi^{(n)}_{t_1t_1t_1t_1} 
 =  6 \, \sum_{r=1}^{n-1} [ (\phi^{(r)}_{t_1} \phi^{(n-r)}_{t_1})_{t_1} - \phi^{(r)}_{t_1} \phi^{(n-r)}_{t_2}
    + \phi^{(r)}_{t_2} \phi^{(n-r)}_{t_1} ] \; .    \label{KPeq-nth}
\ee
Setting
\bez
    \phi^{(n)} = \sum_{i_1,\ldots,i_n=1}^N \frac{\phi_{i_1} \cdots \phi_{i_n}}{(y_{i_1}-x_{i_2})
    \cdots(y_{i_{k-1}}-x_{i_k})}
\eez
with $N \in \mathbb{N}$ (the ``soliton number''), constants $x_i, y_j$, and 
\bez
    \phi_k = c_k \, e^{\xi(\mathbf{t},x_k)} \, e^{- \xi(\mathbf{t},y_k)} \, , \qquad
    \xi(\mathbf{t},x) = \sum_{n \geq 1} t_n \, x^n \, , 
\eez
then (\ref{KPeq-nth}) reduces to
\be
    4 \, \mathbf{p}_1 \mathbf{p}_3 - 3 \, \mathbf{p}_2{}^2 - \mathbf{p}_1{}^4 
 = - 6 \, \mathbf{p}_1 (\mathbf{p}_1 \bu \mathbf{p}_1) 
   + 6 \, ( \mathbf{p}_1 \bu \mathbf{p}_2 - \mathbf{p}_2 \bu \mathbf{p}_1 ) \, ,
        \label{KPeq-id}
\ee
where
\bez
    \mathbf{p}_r = \sum_{k=1}^N (x_k^r-y_k^r)  \qquad \quad   r=1,2,\ldots   \label{p-xy}
\eez
and
\be
  \mathbf{p}_r \bu \mathbf{p}_s 
 = \sum_{1 \leq i< j \leq k \leq N} (x_i^r-y_i^r) \, x_j \, (x_k^s-y_k^s)
   - \sum_{1 \leq i \leq j< k \leq N} (x_i^r-y_i^r) \, y_j \, (x_k^s-y_k^s) \; . \label{pbup}
\ee
By closer inspection, (\ref{KPeq-id}) turns out to be an \emph{identity} for arbitrary $N$ 
(see also section~\ref{sec:KP-ids}).  
It mirrors in an obvious way the structure of the KP equation (\ref{KPeq}) with the correspondence 
expressed by a linear map $\sigma$ such that 
\be
    \sigma(\mathbf{p}_n) = -\phi_{t_n} \, , \qquad
    \sigma(\mathbf{p}_n \, a) = \pa_{t_n}(\sigma(a)) = \sigma(a)_{t_n} \, , \qquad
    \sigma(a \bu b) = \sigma(a) \, \sigma(b) \; .  \label{sigma_Sym}
\ee

Setting $y_k=0$, $k=1,\ldots,N$, (\ref{KPeq-id}) becomes an identity involving quasi-symmetric
polynomials in the ``variables'' $x_k$, $k=1,\ldots,N$. Since this is an identity for any 
$N \in \mathbb{N}$, it is helpful to consider quasi-symmetric functions in an infinite set 
of variables, $x_1,x_2,\ldots$. 
The products in the nonlinear terms of the KP equation then correspond to a new product of 
quasi-symmetric functions, e.g. 
\be
    \Big( \sum_i x_i^r \Big) \bu \Big( \sum_j x_j^s \Big)
  = \sum_{i<j \leq k} x_i^r \, x_j \, x_k^s \; .   \label{KP-product}
\ee
The algebraic structure that emerges in this way, more generally from the KP hierarchy and certain 
extensions, has been elaborated in \cite{DMH05KPalgebra}, but the relation with quasi-symmetric
functions remained unrecognized. It appeared explicitly in a different, though related, approach 
in \cite{DMH06nahier}. 
\vskip.1cm

(\ref{sigma_Sym}) defines the map $\sigma$ on symmetric functions without a constant term, 
and their $\bu$-products. 
Since the product $\bu$ leads out of the algebra of symmetric functions into the bigger algebra 
of quasi-symmetric functions, it is natural to ask for an extension of $\sigma$ to the 
whole algebra $\QSym$. Such an extension indeed has been 
achieved in \cite{DMH05KPalgebra}, and it necessitated the introduction of Moyal-type 
products in the target space (see also \cite{DMH07def}). 
\vskip.1cm

Switching the second set of variables $y_1,y_2,\ldots$ on, 
we are led to a generalization of quasi-symmetric functions to quasi-supersymmetric 
functions, see section~\ref{sec:qss}. The focus of the present work is, however, on the use of the above 
new product in the theory of quasi-symmetric functions, and in particular on relations with 
familiar structures on $\QSym$.

\section{Quasisymmetric functions}
\label{sec:qs}
Let $X$ be a countably infinite \emph{totally ordered} set of commuting variables
and $\mathbb{Z}[[X]]$ the corresponding ring of formal power series 
over $\mathbb{Z}$, which is unital with identity element $1$. 
We denote the ordering relation by $\leq$ and use $<$ for the strict order. 
An element $a$ of $\mathbb{Z}[[X]]$ is a \emph{quasi-symmetric function} if it is of 
bounded degree and if for $x_1<\cdots<x_k$ and $y_1<\cdots<y_k$ in $X$,
and for any choice of positive integers $n_1,\ldots,n_k$, the monomials
$x_1^{n_1}\cdots x_k^{n_k}$ and $y_1^{n_1}\cdots y_k^{n_k}$ have the same
coefficient in $a$ \cite{Gess84,Reut93,Malv+Reut95,Stan99v2}. 
(Our use of $x_i$ and $y_j$ differs from that of section~\ref{sec:KP->qs}.)  
$\QSym$ is a unital subring of $\mathbb{Z}[[X]]$. 
In the following, we consider $\QSym$ as an algebra over $\mathbb{Q}$. 
A basis of $\QSym$ is given by $1$ and
\be
    M_C = \sum_{x_1 < \cdots < x_k} x_1^{n_1} \cdots x_k^{n_k} \, ,  \label{M_C}
\ee
where $C$ denotes the composition $(n_1,\ldots,n_k)$, a sequence of positive integers.  
The sum is over all $x_1,\ldots,x_k \in X$, subject to the ordering condition indicated
under the summation symbol. The weight of compositions supplies $\QSym$ with a natural grading. 
For the above composition $C$, the weight is $|C| = n_1+\cdots +n_k$ and its length is $\ell(C)=k$. 
We set $M_{\emptyset} = 1$, where $\emptyset$ denotes the empty composition. Since 
\bez
      M_{(n)} = \sum_x x^n        \qquad \quad  n = 1,2,\ldots \, , 
\eez
we have
\bez
   M_{(m)} M_{(n)} = \sum_{x_1<x_2} x_1^m x_2^n + \sum_{x_1 \leq x_2} x_1^n x_2^m 
                   = M_{(m,n)} + M_{(m+n)} + M_{(n,m)}  \, , 
\eez
which generalizes to
\be
      M_{(m)} M_{(n_1,\ldots,n_k)} 
  &=& M_{(m,n_1,\ldots,n_k)} + M_{(n_1+m,n_2,\ldots,n_k)} 
    + M_{(n_1,m,n_2,\ldots,n_k)} + M_{(n_1,n_2+m,n_3,\ldots,n_k)} \nonumber \\
  && + \cdots + M_{(n_1,n_2,\ldots,n_k,m)}  \; .
        \label{M_mM_n1...}
\ee
$\QSym$ has a natural Hopf algebra structure \cite{Gess84,Ehre96,Malv+Reut95} with the 
coassociative (outer \cite{Malv+Reut95}) coproduct defined by
\be
    \Delta(M_C) = \sum_{AB=C} M_A \otimes M_B \, ,   \label{Delta(M_C)}
\ee
where the sum is over all compositions $A,B$ that concatenate (with their concatenation denoted by $AB$)
to $C$. The summation includes the empty composition. In particular, 
\bez
    \Delta(1) &=& 1 \otimes 1 \, , \nonumber \\ 
    \Delta(M_{(n)}) &=& 1 \otimes M_{(n)} + M_{(n)} \otimes 1 \, , \nonumber
                      \\
    \Delta(M_{(n_1,n_2)}) &=& 1 \otimes M_{(n_1,n_2)} + M_{(n_1)} \otimes M_{(n_2)} 
                              + M_{(n_1,n_2)} \otimes 1 \; .
\eez
We see that $M_{(n)}$ is a primitive element of the Hopf algebra. 
The counit is determined by $\varepsilon(M_C)=\delta_{C,\emptyset}$. 
We will later meet the antipode $S$. 
$\QSym$ admits another bialgebra structure \cite{Gess84,Malv+Reut95} (with the ``inner coproduct'' 
\cite{Malv+Reut95}), but this will not be considered in this work. 
\vskip.1cm

An alternative basis of $\QSym$ is given by $1$ and 
\bez
  \tilde{M}_{(n_1,\ldots,n_r)} = \sum_{x_1 \leq x_2 \leq \cdots \leq x_r} 
    x_1^{n_1} \cdots x_r^{n_r} \, , 
\eez

For a composition $C = (n_1,\ldots,n_r)$ of $n$, hence $|C| = n$, we define
\bez
    F_C = \sum x_1 \cdots x_n \, ,
\eez
where the summation is over all $x_1,\ldots,x_n \in X$, subject to the conditions
\bez
    x_1 \leq \cdots \leq x_{n_1} < x_{n_1+1} \leq \cdots \leq x_{n_1+n_2} < \cdots
    < x_{n_1+\cdots+n_{r-1}+1} \leq \cdots \leq x_n \; .
\eez
For the empty composition we set $F_{\emptyset} = 1$. 
Then $\{ F_C \, | \, C \mbox{ composition} \}$ 
constitutes the \emph{fundamental basis} of $\QSym$ \cite{Gess84,Stan99v2}. Its elements are 
called \emph{quasi-ribbons} in \cite{GKLLRT95}. 
In particular, we have $F_{(n)} = \tilde{M}_{(1^n)}$ and $F_{(1^n)} = M_{(1^n)}$. 
Expressed in terms of the basis $\{ M_C \}$, we have $F_C = \sum_{D \succeq C} M_D$, 
where the sum is over all compositions $D$ with weight equal to $|C|$ and which are 
finer than $C$ (including $D=C$). 
For example, $F_{(3,1)} = M_{(3,1)} + M_{(2,1,1)} + M_{(1,1,1,1)}$. 
With this notation, we also have $\tilde{M}_C = \sum_{C \succeq D} M_D$.

\section{New products}
\label{sec:new_products}
We introduce a sequence of noncommutative and nonassociative products in $\QSym$ 
($\mathbb{Q}$-linear maps $\QSym \otimes \QSym \rightarrow \QSym$)  
via
\be
       M_A \bu_k M_B 
  &=& \sum_{x_1 < \cdots < x_r < z \leq y_1 < \cdots <y_s} x_1^{n_1} \cdots x_r^{n_r} z^k
       y_1^{m_1}\cdots y_s^{m_s}   \qquad \quad k=1,2,\ldots    \, , 
             \label{M_AbuM_B}
\ee
where $A = (n_1,\ldots,n_r)$ and $B = (m_1,\ldots,m_s)$, and 
\be
    1 \bu_k M_A = \sum_{z \leq x_1 < \cdots < x_r} z^k x_1^{n_1} \cdots x_r^{n_r} \, , \quad
    M_A \bu_k 1 = \sum_{x_1<\cdots <x_r<z} x_1^{n_1}\cdots x_r^{n_r} z^k \, , \quad
    1 \bu_k 1 = \sum_x x^k \; .
           \label{1buM_A_etc}
\ee
For $k=1$, we recover (\ref{KP-product}), i.e. the product that appeared in the above sketched approach to solve the KP equation (and more generally the KP hierarchy). 
With respect to the natural grading of $\QSym$, the product $\bu_k$ determines $\mathbb{Q}$-bilinear maps 
$\QSym^m \otimes \QSym^n \to \QSym^{m+n+k}$. If $G_k$ denotes the semigroup 
$(\mathbb{N} \cup \{0\}, +_k)$, where $i+_k j = i+j+k$, then $(\QSym,\bu_k)$ is a 
$G_k$-graded algebra.
As a consequence of the above definitions, we have
\bez
    M_A \bu_n 1 = M_{A(n)} \, , 
\eez
and thus
\be
         M_{(n_1,\ldots,n_k)} 
  = M_{(n_1,\ldots,n_{k-1})} \bu_{n_k} 1  
  = (\cdots((1\bu_{n_1} 1) \bu_{n_2}1) \bu_{n_3} \cdots) \bu_{n_k}1 \; . \label{M_n-iter}
\ee
Furthermore, 
\be
    M_A \bu_n M_{(m)B} = M_{A(n,m)B} + M_{A(n+m)B} \, ,  
                           \label{M_Abu_nM_mB}
\ee
where $A,B$ may be empty. 
We will also use the notation
\bez
    a \bu_k b = \mathbf{m}_k (a \otimes b)   \qquad \quad \forall a,b \in \QSym \; .
\eez
Restricted to the non-unital ring $\QSym'$, obtained from $\QSym$ 
without the constant elements, i.e. $\QSym' = \QSym/\mathbb{Q}1$, 
the new products are associative and also combined associative. 
But in general they are \emph{not} associative. Indeed, we have the following property, 
which also shows that the products $\bu_k$, $k>1$, can all be expressed in terms of the 
first product $\bu = \bu_1$.

\begin{lemma} 
\label{lemma:bu_k-iter}
For all $a,b \in \QSym$ and $k,l=1,2,\ldots$ we have
\be
    a \bu_k (1 \bu_l b) - (a \bu_k 1) \bu_l b = a \bu_{k+l} b \; .  \label{bu_k-iter}
\ee
\end{lemma}
\begin{proof}
By linearity it is sufficient to consider (\ref{bu_k-iter}) for the 
elements of the basis $\{M_C \, | \, C \mbox{ composition} \}$. But for the latter, 
(\ref{bu_k-iter}) is immediately verified by use of the definitions (\ref{M_C}), 
(\ref{M_AbuM_B}) and (\ref{1buM_A_etc}). 
\end{proof}

\begin{proposition}
\label{prop:QSym-free}
$\QSym$ is generated by $1$ and the nonassociative product $\bu$. 
\end{proposition}
\begin{proof}
Since $1$ and the elements $M_{(n_1,\ldots,n_k)}$ 
constitute a basis of $\QSym$, the assertion follows from (\ref{M_n-iter}) and lemma~\ref{lemma:bu_k-iter}. 
\end{proof}

The following is also easily verified. 

\begin{proposition}
$\QSym$, supplied with any of the products $\bu_k$, $k \in \mathbb{N}$, is a 
weakly nonassociative algebra. More generally, we have
\be
  ( a \bu_k (b \bu_m c) ) \bu_n d = a \bu_k ( (b \bu_m c) \bu_n d ) \label{bu_k-wna}
\ee
for all $a,b,c,d \in \QSym$ and all $k,m,n = 1,2,\ldots$. 
\hfill $\square$
\end{proposition}
\vskip.1cm

The alternative basis elements $\tilde{M}_C$ of $\QSym$ are recursively determined by 
\be
    \tilde{M}_{(n)C} = 1 \bu_n \tilde{M}_C \; .  \label{tM_(n)C}
\ee
Furthermore, for any two compositions $A,B$, we have
\be
    \tilde{M}_{A(m)} \bu_n \tilde{M}_B = \tilde{M}_{A(m,n)B} - \tilde{M}_{A(m+n)B} \; .
         \label{tM_Ambu_ntM_B}
\ee
\vskip.1cm

In case of the fundamental basis elements, it is quite easily verified that
\be
       F_A \bu F_{(m)B} = F_{A(m+1)B}   \label{F_C-factorization-wrt-bu}
\ee
for any two compositions $A,B$, and $m=1,2,\ldots$. 
Of particular importance are compositions of the form $(m+1, 1^n)$, 
$m,n=0,1,\ldots$, where $(1^n) = (1,\ldots,1)$ (with $n$ entries), for which we find
\be
    F_{(m+1,1^n)} = \mathcal{L}_1^m \mathcal{R}_1^n (1 \bu 1) \, ,  \label{F-LR(1bu1)}
\ee
where $\mathcal{L}_a b = a \bu b$ and $\mathcal{R}_a b = b \bu a$. 
Note that on the right hand side of (\ref{F-LR(1bu1)}) the left and right multiplications  
commute as a consequence of the weak nonassociativity property (\ref{weak_nonassoc}) of 
the product $\bu$. 
Since any non-empty composition $C$ that is not of the form $(m+1,1^n)$ can be written as 
$C=(m_1+1, 1^{n_1}, m_2 + 2, 1^{n_2}, \ldots, m_r +2, 1^{n_r})$ 
with $m_1,n_1,\ldots,m_r,n_r \in \mathbb{N} \cup \{0\}$, 
as a consequence of (\ref{F_C-factorization-wrt-bu}) we have
\be
   F_C = F_{(m_1+1,1^{n_1})} \bu F_{(m_2+1,1^{n_2})} \bu \cdots \bu F_{(m_r+1,1^{n_r})} \, , 
         \label{F_factoriz}
\ee
where the right hand side has an associative structure due to (\ref{weak_nonassoc}). 
Hence, $F_C$ factorizes into a $\bu$-product of elementary basis elements 
(\ref{F-LR(1bu1)}).

\begin{remark}
Instead of setting the $y$'s to zero in (\ref{pbup}), we may set the 
$x$'s to zero. With a change of sign, this leads to the alternative 
weakly nonassociative products in $\QSym$ determined by 
$1 \hat{\bu}_k 1 = \sum_x x^k$, 
\bez
 && M_A \, \hat{\bu}_k \, M_B = \sum_{x_1 < \cdots < x_r \leq z < y_1 < \cdots <y_s} x_1^{n_1} 
       \cdots x_r^{n_r} z^k y_1^{m_1}\cdots y_s^{m_s}  \nonumber \\
 && 1 \hat{\bu}_k M_A = \sum_{z < x_1 < \cdots < x_r} z^k x_1^{n_1} \cdots x_r^{n_r} \, , \quad
    M_A \hat{\bu}_k 1 = \sum_{x_1<\cdots <x_r \leq z} x_1^{n_1}\cdots x_r^{n_r} z^k 
    \, , \qquad
\eez
for which all results in this work have a counterpart. In particular, we obtain  
$F_{(1^m,n+1)} = \hat{\mathcal{L}}_1^m \hat{\mathcal{R}}_1^n (1 \hat{\bu} 1)$ and 
any $F_C$ with $C$ not of the form $(1^m,n+1)$ 
factorizes into a $\hat{\bu}$-product of such elementary elements. 
\end{remark}

\section{Relations between the old and the new products, and the coproduct}
\label{sec:relations}
The following result allows us to express the ordinary product of quasi-symmetric functions 
recursively in terms of the product $\bu$.

\begin{proposition}
For any composition $C$, any $a \in \QSym$, and $k=1,2,\ldots$,
\be
     M_{C(k)} \, a = \mathbf{m}_k \Big( (M_C \otimes 1) \, \Delta(a) \Big)   \; .
          \label{M_Cma}
\ee
\end{proposition}
\begin{proof}
With $D=(m_1,\ldots,m_s)$ and $E=(n_1,\ldots,n_r)$, 
\bez
     M_D M_E = \left( \sum_{y_1< \cdots <y_s} y_1^{m_1} \cdots y_s^{m_s} \right)
        \left(\sum_{x_1< \cdots <x_r} x_1^{n_1} \cdots x_r^{n_r} \right)
\eez
can be written as 
\bez
   \sum_{0 \leq i_1 \leq \cdots \leq i_s \leq r} \sum 
   x_1^{n_1} \cdots x_{i_1}^{n_{i_1}}y_1^{m_1}x_{i_1+1}^{n_{i_1+1}} \cdots
   x_{i_{s-1}}^{n_{i_{s-1}}}y_{s-1}^{m_{s-1}}x_{i_{s-1}+1}^{n_{i_{s-1}+1}} \cdots
   x_{i_s}^{n_{i_s}}y_s^{m_s}x_{i_s+1}^{n_{i_s+1}} \cdots x_r^{n_r} \, , 
\eez
where any expression of the form $x_{k}^{n_{k}} \cdots x_{l}^{n_{l}}$ 
should be replaced by $1$ if $k > l$. The inner summation is over all 
$x_1,\ldots,x_r, y_1,\ldots,y_s\in X$, subject to the condition
\bez
    x_1< \cdots < x_{i_1}<y_1 \leq x_{i_1+1}< \cdots <x_{i_{s-1}}<y_{s-1} \leq x_{i_{s-1}+1}
    < \cdots <x_{i_s}<y_s \leq x_{i_s+1}< \cdots <x_r \; .
\eez
For fixed $i_s$ is
\bez
   \sum_{0 \leq i_1 \leq \cdots \leq i_{s-1} \leq i_s} \sum 
   x_1^{n_1} \cdots x_{i_1}^{n_{i_1}}y_1^{m_1}x_{i_1+1}^{n_{i_1+1}} \cdots
   x_{i_{s-1}}^{n_{i_{s-1}}}y_{s-1}^{m_{s-1}}x_{i_{s-1}+1}^{n_{i_{s-1}+1}} \cdots
   x_{i_s}^{n_{i_s}}y_s^{m_s}x_{i_s+1}^{n_{i_s+1}} \cdots x_r^{n_r}
\eez
equal to $(M_{D'} M_A) \bu_{m_s} M_B$, where $A=(n_1,\ldots, n_{i_s})$, $B=(n_{i_s+1},\ldots, n_r)$
and $D'=(m_1,\ldots,m_{s-1})$. Summing over $0 \leq i_s \leq r$, we thus obtain
\bez
    M_D M_E = \sum_{AB=E} (M_{D'} M_A) \bu_{m_s} M_B \; .
\eez
\end{proof}

In particular, (\ref{M_mM_n1...}) can be expressed as 
\bez
   M_{(m)} M_{(n_1,\ldots,n_k)} &=& \mathbf{m}_m \circ \Delta(M_{(n_1,\ldots,n_k)}) \nonumber \\
  &=& 1 \bu_m M_{(n_1,\ldots,n_k)} + M_{(n_1)} \bu_m M_{(n_2,\ldots,n_k)} + \cdots 
   + M_{(n_1,\ldots,n_k)} \bu_m 1  \; .
\eez

Next we define a bimodule structure on $\QSym \otimes \QSym$ via
\bez
    (a \otimes b) \bu_k c = a \otimes (b \bu_k c) \, , \qquad
    c \bu_k (a \otimes b) = (c \bu_k a) \otimes b \, , 
\eez
for all $a,b,c \in \QSym$. As a consequence, the usual bimodule property holds, i.e. 
\bez
     a \bu_k (\mathfrak{m} \bu_l b) = (a \bu_k \mathfrak{m}) \bu_l b 
       \qquad \quad \forall a,b \in \QSym \, , \quad  \forall \mathfrak{m} \in \QSym \otimes \QSym \; .
\eez
But we have the following restricted form of the usual left and right module properties 
(see also \cite{DMH06nahier}), 
\bez
   a \bu_k (b \bu_l \mathfrak{m}) = (a \bu_k b) \bu_l \mathfrak{m} \, , \quad
  (\mathfrak{m} \bu_k b) \bu_l a = \mathfrak{m} \bu_k (b \bu_l a) \qquad \quad 
      \forall b \in \QSym' \, , 
\eez
whereas 
\bez
   a \bu_k (1 \bu_l \mathfrak{m}) = (a \bu_k 1) \bu_l \mathfrak{m} + a \bu_{k+l} \mathfrak{m} \, , 
   \qquad
   (\mathfrak{m} \bu_k 1) \bu_l a = \mathfrak{m} \bu_k (1 \bu_l a) - \mathfrak{m} \bu_{k+l} a \; .
\eez
\vskip.1cm

The following proposition turns $(\QSym,\bu, \Delta)$ into an \emph{infinitesimal bialgebra}
\cite{Joni+Rota79,Hirs+Raph92,Agui00,Agui01,Agui02,Agui04,Agui+Loday04,Ehre+Read02}.\footnote{This 
has to be distinguished from a ``unital infinitesimal bialgebra'' as defined in  
\cite{Loday+Ronco06} (see also \cite{Leroux04,Foissy08}). 
}
We should stress, however, that here $(\QSym,\bu)$ is \emph{not} associative, so that we 
are not quite in the framework of the latter references. 
Furthermore, we note that $(\QSym,\bu)$ is \emph{not} unital. In fact, any infinitesimal 
bialgebra possessing a unit and a counit is trivial \cite{Agui00}.

\begin{proposition}
\label{prop:Delta-deriv}
The coproduct of $\QSym$ acts as a derivation on the products $\bu_n$, i.e.
\be
    \Delta(a \bu_n b) = \Delta(a) \bu_n b + a \bu_n \Delta(b) \qquad \quad n=1,2,\ldots \, ,
         \label{Delta-bu-derivation}
\ee
for all $a,b \in \QSym$.
\end{proposition}
\begin{proof}
We have
\bez
     \Delta(M_A) \bu_n 1 + M_A \bu_n \Delta(1) 
 &=& \sum_{CD=A} M_C\otimes (M_D \bu_n 1) + (M_A \bu_n 1) \otimes 1 \\
 &=& \sum_{CD=A} M_C \otimes M_{D(n)}  + M_{A(n)} \otimes 1  \\
 &=& \sum_{CD=A(n)} M_C \otimes M_D 
  = \Delta(M_{A(n)})
  = \Delta(M_A \bu_n 1)  \, ,
\eez
and also
\bez
    \lefteqn{ \Delta(M_A) \bu_n M_{(m)B} + M_A \bu_n \Delta(M_{(m)B}) } \hspace*{1cm}&& \\
 &=& \sum_{CD=A} M_C \otimes (M_D \bu_n M_{(m)B}) + \sum_{CD=(m)B} (M_A \bu_n M_C) \otimes M_D \\
 &=& \sum_{CD=A} M_C \otimes M_{D(n,m)B} + \sum_{CD=A} M_C \otimes M_{D(n+m)B} \\
 &&  + M_{A(n)} \otimes M_{(m)B} + \sum_{CD=B} M_{A(n,m)C} \otimes M_D 
     + \sum_{CD=B} M_{A(n+m)C} \otimes M_D \\
 &=& \sum_{CD=A(n,m)B} M_C \otimes M_D + \sum_{CD=A(n+m)B} M_C \otimes M_D 
  = \Delta(M_A \bu_n M_{(m)B}) \; .
\eez
\end{proof}

Using (\ref{tM_(n)C}) and (\ref{Delta-bu-derivation}), one easily proves by induction that
\be
     \Delta(\tilde{M}_C) = \sum_{AB=C} \tilde{M}_A \otimes \tilde{M}_B  \; . \label{Delta(tM_C)}
\ee

\begin{proposition}
\label{prop:distr}
The distributivity rule  
\be
    c \, (a \bu_m b) = \mathbf{m}_m \left( \Delta(c) \, (a \otimes b) \right) 
    \label{distr}
\ee
holds for all $a,b,c \in \QSym$ and $m=1,2,\ldots$. 
\end{proposition}
\begin{proof}
Clearly, the assertion is true for $c = M_{\emptyset} =1$. 
An application of (\ref{M_Cma}) and (\ref{Delta-bu-derivation}) yields
\bez
     M_{C(n)} (a \bu_m b) 
 &=& \mathbf{m}_n \left( (M_C \otimes 1) \, \Delta(a \bu_m b) \right) \\
 &=& \mathbf{m}_n \Big( (M_C \otimes 1) \, \left( \Delta(a) \bu_m b + a \bu_m \Delta(b) \right) \Big) \\
 &=& \mathbf{m}_n \Big( (M_C \otimes 1) \Big( \sum_{(a)}  \, a_{[1]}  \otimes a_{[2]} \bu_m b 
     + \sum_{(b)} a \bu_m b_{[1]} \otimes b_{[2]} \Big) \Big) \\
 &=& \mathbf{m}_n \Big( \sum_{(a)} M_C a_{[1]} \otimes a_{[2]} \bu_m b 
     + \sum_{(b)} M_C (a \bu_m b_{[1]}) \otimes b_{[2]} \Big) \\
 &=& \sum_{(a)} (M_C a_{[1]}) \bu_n (a_{[2]} \bu_m b) 
      + \sum_{(b)} ( M_C (a \bu_m b_{[1]}) ) \bu_n b_{[2]} \; .
\eez
Here we used the Sweedler notation $\Delta(a) = \sum_{(a)} a_{[1]}  \otimes a_{[2]}$.  
We proceed by induction and assume that the assertion holds for $c=M_C$ with $\ell(C)=r$. 
Hence
\bez
    M_{C(n)} (a \bu_m b) 
  = \sum_{(a)} (M_C a_{[1]}) \bu_n (a_{[2]} \bu_m b)
   + \sum_{(b)} \Big( \sum_{AB=C} (M_A \, a) \bu_m (M_B \, b_{[1]}) \Big) \bu_n b_{[2]} \, ,
\eez
by use of (\ref{Delta(M_C)}). The first term on the right hand side of the last equation can then 
be rewritten as follows, 
\bez
    \sum_{(a)} (M_C a_{[1]}) \bu_n (a_{[2]} \bu_m b)
  = {\sum_{(a)}}' (M_C \, a_{[1]}) \bu_n a_{[2]} \bu_m b + (M_C \, a) \bu_n (1 \bu_m b) \, ,
\eez
where the primed summation omits the term involving the summand $a \otimes 1$ of $\Delta(a)$, 
which (by recalling (\ref{bu_k-wna})) is the only term with a \emph{non}associative 
structure (now the last term on the right hand side). 
Similarly, the second term on the right hand side of the previous equation can be written as
\bez
   \sum_{(b),AB=C} ((M_A \, a) \bu_m (M_B \, b_{[1]})) \bu_n b_{[2]}
 = {\sum_{(b),AB=C}}' (M_A \, a) \bu_m (M_B \, b_{[1]}) \bu_n b_{[2]} 
   + ((M_C \, a) \bu_m 1) \bu_n b \, ,
\eez
where the primed summation omits the only summand (corresponding to the summand 
$1 \otimes b$ of $\Delta(b)$ and $B=\emptyset$) that has a nonassociative structure. 
Using 
\bez
    (M_C \, a) \bu_n (1 \bu_m b) + ((M_C \, a) \bu_m 1) \bu_n b
  = ((M_C \, a) \bu_n 1) \bu_m b + (M_C \, a) \bu_m (1 \bu_n b) \, , 
\eez
which is an immediate consequence of (\ref{bu_k-iter}), we find that
\bez
    M_{C(n)} (a \bu_m b) = \sum_{(a)} ((M_C \, a_{[1]}) \bu_n a_{[2]}) \bu_m b
  + \sum_{(b),AB=C}(M_A \, a) \bu_m ((M_B \, b_{[1]}) \bu_n b_{[2]}) \; .
\eez
With the help of (\ref{M_Cma}), this becomes
\bez
   M_{C(n)} (a \bu_m b) 
 = (M_{C(n)} \, a) \bu_m b + \sum_{AB=C} (M_A \, a) \bu_m (M_{B(n)} \, b)
 = \sum_{AB=C(n)} (M_A \, a) \bu_m (M_B \, b) \, ,
\eez
which completes the induction step.
\end{proof}

\section{Relation between the antipode and the new products}
\label{sec:antipode} 
Let us define a linear map $S \, : \, \QSym \rightarrow \QSym$ by 
\be
    S(M_C) = (-1)^{\ell(C)} \, \tilde{M}_{\tilde{C}} \, ,  \label{antipode-formula}
\ee
where $\tilde{C}$ is the reverse of the composition $C$, i.e. $C$ written in 
reversed order. In particular, $S(1)=1$.

\begin{proposition}
\label{prop:S-bu}
\be
    S(a \bu_n b) = - S(b) \bu_n S(a)  \qquad \quad \forall a,b \in \QSym \, , 
                   \quad n=1,2,\ldots  \; .
                   \label{S(bu)}
\ee
\end{proposition}
\begin{proof}
\bez
    S(M_C \bu_n 1) = S(M_{C(n)})
                   = - (-1)^{\ell(C)} \tilde{M}_{(n) \tilde{C}}
                   = - (-1)^{\ell(C)} \, 1 \bu_n \tilde{M}_{\tilde{C}}
                   = - S(1) \bu_n S(M_C) \; .
\eez
Furthermore, we have
\bez
     S(M_A \bu_n M_{(m)B}) 
 &=& S(M_{A(n,m)B} + M_{A(n+m)B})
  = (-1)^{\ell(A)+\ell(B)} ( \tilde{M}_{\tilde{B}(m,n)\tilde{A}}
                            -\tilde{M}_{\tilde{B} (n+m) \tilde{A}}) \\
 &=& (-1)^{\ell(A)+\ell(B)} \tilde{M}_{\tilde{B}(m)} \bu_n \tilde{M}_{\tilde{A}}
  = - S(M_{(m)B}) \bu_n S(M_A)  \, , 
\eez
where we used (\ref{M_Abu_nM_mB}) and (\ref{tM_Ambu_ntM_B}).
\end{proof}

Next we prove that $S$ is the antipode of the Hopf algebra of quasi-symmetric functions.

\begin{proposition}
If $\mu$ denotes the usual product of quasi-symmetric functions, and $\varepsilon$ the counit, 
then 
\be
   \mu \, (\mathrm{id} \otimes S) \, \Delta = 1 \, \varepsilon 
 = \mu \, (S \otimes \mathrm{id}) \, \Delta  \; .  \label{antipode-conds}
\ee
\end{proposition}
\begin{proof}
Since $\mu \, (\mathrm{id} \otimes S) \, \Delta(1) = 1 
= \mu \, (S \otimes \mathrm{id}) \, \Delta(1)$, it is sufficient to verify that both sides 
of (\ref{antipode-conds}) applied to $M_C$ vanish if $C$ is not empty. Hence we have to 
show that 
\bez
    \sum_{DE=C} M_D \, S(M_E) = 0  \qquad \mbox{and} \qquad
    \sum_{DE=C} S(M_D) \, M_E = 0 \, , 
\eez
for all $C$ different from the empty composition. 
We concentrate on the first relation and use induction on the length $\ell(C)$ of the 
composition $C$. For a composition of length $1$, we obtain
\bez
    \sum_{DE=(n)} M_D \, S(M_E) 
 &=& 1 \, S(M_{(n)}) + M_{(n)} \, 1 
  = S(M_{(n)}) + M_{(n)} = S(1 \bu_n 1) + M_{(n)} \\
 &=& - S(1) \bu_n S(1) + M_{(n)} 
  = - 1 \bu_n 1 + M_{(n)} 
  = - M_{(n)} + M_{(n)} = 0 \; .
\eez
Here we used (\ref{S(bu)}). 
Let us now assume that the assertion holds for all $C$ with $\ell(C) \leq r$. Then we have
\bez
    & &   \sum_{DE=C(n)} M_D \, S(M_E) - M_{C(n)}
     = \sum_{DE=C} M_D \, S(M_E \bu_n 1) 
     = - \sum_{DE=C} M_D \, (1 \bu_n S(M_E)) \\
    &=& - \sum_{A B E=C} M_A \bu_n (M_B \, S(M_E)) 
     = - \sum_{A D = C} M_A \bu_n \Big( \sum_{B E=D} M_B \, S(M_E) \Big) \; .
\eez
Besides (\ref{S(bu)}), we applied (\ref{distr}). 
By induction hypothesis, $\sum_{BE=D} M_B \, S(M_E)$ vanishes, except for the case where 
$D$ is the empty composition, where this is equal to $1$. Hence 
\bez
    \sum_{DE=C(n)} M_D \, S(M_E) = M_{C(n)} - M_C \bu_n 1 = 0 \; .
\eez
The second relation in (\ref{antipode-conds}) can be proved in the same way. 
\end{proof}

Formula (\ref{antipode-formula}) for the antipode appeared in \cite{Ehre96,Malv+Reut95}. 
As a side-result, we obtained a new proof of this expression for the antipode. 
Since $\QSym$ is commutative with respect to the original product, the 
antipode satisfies $S^2 = \mathrm{id}$ (see e.g. \cite{Klim+Schm97}, p.15). 
We should mention that $S$ is \emph{not} an antipode of an \emph{infinitesimal 
Hopf algebra} as defined in \cite{Agui00}. 
\vskip.1cm

(\ref{S(bu)}) shows that (up to a sign) the antipode exchanges left and 
right multiplication with the product $\bu$. This implies 
\bez
    S(F_{(m+1,1^n)}) = (-1)^{m+1+n} \, F_{(n+1,1^m)}
\eez
for $m,n=0,1,\ldots$. 
Using (\ref{F_factoriz}) and (\ref{S(bu)}), we quickly recover the following 
result (see \cite{Ehre96,Malv+Reut95}), 
\bez
     S(F_C) = (-1)^{|C|} \, F_{\omega(C)} \; .  
\eez
The simple calculation determines the map $\omega$ of compositions as follows.  
If $C=(m+1,1^n)$, then $\omega(C) = (n+1,1^m)$. If a composition $C$ is not of this form, 
then it can be written uniquely as $C = (m_1+1, 1^{n_1}, m_2 + 2, 1^{n_2}, \ldots, m_r +2, 1^{n_r})$ 
with $m_1,n_1,\ldots,m_r,n_r \in \mathbb{N} \cup \{0\}$ and $r>1$, and we set 
$\omega(C) = (n_r+1,1^{m_r},n_{r-1}+2,1^{m_{r-1}},\ldots,n_1+2,1^{m_1})$. 
Note that $\omega^2 = \mathrm{id}$.

\section{KP identities}
\label{sec:KP-ids}
We recall that the symmetric functions form a subalgebra $\Sym$ of $\QSym$, and a basis is given by 
products of the (homogeneous) \emph{complete symmetric functions} 
\bez
    h_n = \tilde{M}_{(1^n)}  \qquad \mbox{where}  \qquad
    (1^n) = ( \underbrace{ 1, \ldots, 1 }_{\mbox{$n$ times}} ) \, , 
\eez
including $h_0 =1$. In the following, we write
\bez
     p_n = M_{(n)}  \qquad \quad n=1,2,\ldots \, , 
\eez
which is the $n$-th power sum. By use of Newton's identities
\be
 n \, h_n = \sum_{k=1}^n p_k \, h_{n-k}   \qquad\quad n=1,2,\ldots \, , 
        \label{h_n-p_k}
\ee
$h_n$, $n>0$, can be recursively expressed in terms of $p_k$, $k=1,\ldots,n$.

\begin{remark}
Introducing the formal sums
\bez
    h(\zeta) = \sum_{n \geq 0} \zeta^n \, h_n \, , \qquad
    p(\zeta) = \sum_{n \geq 1} \zeta^{n-1} p_n \, , 
\eez
with an indeterminate $\zeta$, (\ref{h_n-p_k}) leads to
\bez
    \frac{\mathrm{d}}{\mathrm{d} \zeta} h(\zeta) = p(\zeta) \, h(\zeta) \, , 
\eez
which integrates to 
\bez
     h(\zeta) = \exp\Big( \sum_{n \geq 1} \frac{\zeta^n}{n} \, p_n \Big) \; .
\eez
It follows that $h_n$ can be expressed as 
\bez
     h_n = \mathbf{s}_n(p_1, p_2/2, p_3/3,\ldots)
\eez
in terms of the elementary Schur polynomial $\mathbf{s}_n$. 
\end{remark}

As a preparation for the main result of this section, we recall the divided power structure 
of the coproduct of $h_n$,
\be
    \Delta(h_n) = \sum_{k=0}^n h_{k} \otimes h_{n-k}  \qquad\quad  n=1,2,\ldots \, ,
        \label{Delta(h_n)}
\ee
which is a special case of (\ref{Delta(tM_C)}).

\begin{proposition}
For $m,n=1,2,\ldots$, we have the following identities,
\be
     h_m \, h_{n+1} - h_{m+1} \, h_n 
  = \sum_{k=1}^m h_k \bu (h_{m-k} \, h_n) - \sum_{k=1}^n h_k \bu (h_{n-k} \, h_m) \; .
      \label{KPids}
\ee
\end{proposition}
\begin{proof}
Using (\ref{distr}) and (\ref{Delta(h_n)}), we find 
\bez
    h_m \, h_{n+1} 
 &=& h_m \, (1 \bu h_n) 
 = \mathbf{m} ( \Delta(h_m) \, (1 \otimes h_n) )
 = \sum_{k=0}^m \mathbf{m} ( (h_k \otimes h_{m-k})(1 \otimes h_n) ) \\
 &=& \sum_{k=0}^m h_k \bu (h_{m-k} h_n ) \; .
\eez
This implies (\ref{KPids}).
\end{proof}

(\ref{KPids}) is a sequence of identities for \emph{symmetric} functions, but 
in the space of \emph{quasi-symmetric} functions, since the product $\bu$ leads outside the 
subspace of symmetric functions. 
For $m=1,n=2$, we obtain
\bez
    h_1 \, h_3 - h_2 \, h_2 = h_1 \bu h_2 - h_1 \bu h_1^2 - h_2 \bu h_1 \; .
\eez
Expressed in terms of $p_n$ via (\ref{h_n-p_k}), this takes the form
\be
   4 \, p_1 \, p_3 - 3 \, p_2^2 - p_1^4 
= - 6 \, p_1 \, ( p_1 \bu p_1 )
  + 6 \, ( p_1 \bu p_2 - p_2 \bu p_1 ) \, ,
    \label{KPid}
\ee
where we used (\ref{distr}) to write $p_1^2 \bu p_1 + p_1 \bu p_1^2 = p_1 \, ( p_1 \bu p_1 )$. 
We observe that the identity (\ref{KPid}) is the KP identity (\ref{KPeq-id}) (for vanishing $y_1,y_2,\ldots$ 
and $N=\infty$). 
We already explained in section~\ref{sec:KP->qs} how the KP equation can be reconstructed from this identity. 
For a partition $\lambda = (\lambda_1,\lambda_2, \ldots , \lambda_r)$, let
\be
    p_\lambda = p_{\lambda_1} \, p_{\lambda_2} \ldots p_{\lambda_r} \; . \label{p_lambda}
\ee
Then $\sigma(p_\lambda) = -\phi_{t_{\lambda_1} t_{\lambda_2} \ldots t_{\lambda_r}}$. 
Since the symmetric functions (\ref{p_lambda}) form a basis of $\Sym$ (over $\mathbb{Q}$) \cite{Macd95}, it 
follows that to any symmetric function $f \in \Sym/\mathbb{Q}1$ there corresponds an expression 
$\sigma(f) = -F \, \phi$ with a differential operator $F(\pa_{t_1}, \pa_{t_2}, \ldots)$, having coefficients 
in $\mathbb{Q}$ and no term of $0$-th order. Furthermore, 
$\sigma(f_1 \bu f_2 \bu \cdots \bu f_r) = (-F_1 \phi) \cdots (-F_r \phi)$ for $f_i \in \Sym/\mathbb{Q}1$, 
$i=1,\ldots,r$. Applying $\sigma$ to (\ref{KPids}), one recovers the whole (noncommutative) KP hierarchy. 
A formulation of the KP hierarchy that corresponds to (\ref{KPids}) in this way, 
can be found e.g. in \cite{DNS89,DMH06func}. 
Of course, one can apply the procedure in section~\ref{sec:KP->qs} of solving the KP equation more generally 
to any member of the KP hierarchy. The fact that (\ref{KPids}) is a sequence of identities would then prove 
that the method indeed generates solutions of the whole KP hierarchy. 
\vskip.1cm

We conjecture that \emph{any} identity in $\QSym$, that is built from symmetric functions 
in $\Sym/\mathbb{Q}1$ and \emph{only} the product $\bu$, corresponds to a partial differential 
equation that is satisfied as a consequence of the KP hierarchy. 
\vskip.1cm

There is another way to describe the correspondence between the identities (\ref{KPids}) 
and the equations of the KP hierarchy. 
According to (\ref{distr}), the primitive element $p_n = M_{(n)}$ acts on a product $a \bu_k b$ as a 
\emph{derivation}. Hence
\bez
   \delta_n(a) = \mathbf{m}_n \circ \Delta(a) = p_n \, a   \qquad  \quad \forall a \in \QSym
\eez
defines a sequence of commuting derivations on $\QSym$ with respect to (any of) the products $\bu_k$, 
i.e.
\bez
   \delta_n(a \bu_k b) = \delta_n(a) \bu_k b + a \bu_k \delta_n(a) \, , \qquad 
   \delta_m \delta_n = \delta_n \delta_m \; .
\eez
This makes contact with the framework developed in \cite{DMH06nahier,DMH08isl} for 
weakly nonassociative algebras. 
Expressing (\ref{KPid}) in terms of these derivations, we obtain
\bez
    4 \, \delta_1 \delta_3(1) - 3 \, \delta_2^2(1) - \delta_1^4(1)
 = - 6 \, \delta_1 (\delta_1(1) \bu \delta_1(1)) 
   + 6 \, ( \delta_1(1) \bu \delta_2(1) - \delta_2(1) \bu \delta_1(1) ) \, ,
\eez
which becomes the KP equation (\ref{KPeq}) via 
$1 \mapsto -\phi$, $\delta_n \mapsto \pa_{t_n}$, and with $\bu$ replaced by the product in 
the associative algebra where $\phi$ takes its values. 
This correspondence extends to the whole KP hierarchy \cite{DMH06nahier,DMH08isl}.

\section{Quasi-supersymmetric functions}
\label{sec:qss}
(\ref{pbup}) shows that the product corresponding to the nonlinearities
of the KP equation involves \emph{two} sets of parameters, the $x$'s and the
$y$'s. Using the usual product, the expressions $\mathbf{p}_r$ defined in (\ref{p-xy}) 
with $N=\infty$ generate \emph{supersymmetric} functions \cite{Bala+Bars81,Scheu84,Stem85,Nimm88,Nimm89} 
(called ``bisymmetric functions'' in \cite{MNR81}), see also \cite{Macd95} 
(example 23 of section I.3). 
Hence we should expect to encounter more generally ``quasi-supersymmetric functions''. 
Such a generalization of supersymmetric functions has not yet appeared in the literature, 
according to our knowledge.  

Let $\mathbf{x} = (x_1,x_2,\ldots)$ and $\mathbf{y} = (y_1,y_2,\ldots)$ be two countably infinite 
sequences of commuting variables, and let $\mathbb{Q}[[\mathbf{x},\mathbf{y}]]$ be the algebra of 
formal power series in the latter with coefficients in $\mathbb{Q}$.
For any monomial of bounded degree, $a = z_{i_1}^{n_1}\cdots z_{i_r}^{n_r}$ where $z_i$ is either $x_i$ 
or $y_i$, let $m(a)$ and $M(a)$ denote the minimal respectively maximal element of $\{i_1,\ldots,i_r\}$.
We extend the previously defined products by setting 
\bez
    1 \bu_n 1 &=& \sum_i (x_i^n -y_i^n) \, , \nonumber \\
    1 \bu_n a &=& \sum_{i \leq m(a)} x_i^n \, a - \sum_{i<m(a)} y_i^n \, a \, , \nonumber \\
    a \bu_n 1 &=& \sum_{M(a)<i} a \, x_i^n - \sum_{M(a) \leq i} a \, y_i^n \, , \nonumber \\
    a \bu_n b &=& \sum_{M(a)<i \leq m(b)} a \, x_i^n b - \sum_{M(a) \leq i < m(b)} a \, y_i^n \, b \, ,
\eez
for monomials $a,b$ of bounded degree. 
Here a sum contributes zero if there is no index $i$ satisfying the conditions underneath the 
respective summation symbol.
These definitions extend to the whole of $\mathbb{Q}[[\mathbf{x},\mathbf{y}]]$ by linearity.
Again, this defines weakly nonassociative products that satisfy (\ref{bu_k-iter}) and (\ref{bu_k-wna}). 
Also in this case all products can be expressed in terms of the first. 
Based on further developments of the theory of weakly nonassociative algebras, we will show in a separate work 
that the weakly nonassociative subalgebra of $\mathbb{Q}[[\mathbf{x},\mathbf{y}]]$, 
generated by $1$ via $\bu$, is closed under the usual multiplication, and that it contains the supersymmetric 
functions. A substitution $x_i=t$ and $y_i=t$ for the \emph{same} $i$ results in expressions 
independent of $t$ (a central property of supersymmetric functions \cite{Scheu84,Stem85}). 
Moreover, the space of quasi-supersymmetric functions is spanned by 
$M_\emptyset = 1$ and the elements defined recursively by 
\bez
    M_{C(n)} = M_C \bu_n 1 \, , 
\eez
for any composition $C$. By use of these results, one derives a sequence of KP identities in 
the algebra of quasi-supersymmetric functions, which are in one-to-one correspondence with identities 
derived in section~\ref{sec:KP-ids} and of which (\ref{KPeq-id}) is the simplest (non-trivial).

\section{Final remarks}
\label{sec:final} 
We supplied the algebra of quasi-symmetric functions with a weakly nonassociative 
product and studied its relations with the ordinary product and the coproduct. 
The new product $\bu$ turned out to be a useful tool in the theory of 
quasi-symmetric functions, despite of the fact that it introduces a weak form 
of nonassociativity. It should be of interest to study more generally 
weakly nonassociative algebras that admit an infinitesimal coproduct, and in this way to 
extend the results in \cite{DMH06nahier,DMH08isl}. 
Such a generalization will be elaborated in a separate work, including an exploration 
of the algebra of quasi-supersymmetric functions. 

The Hopf algebra of quasi-symmetric functions is the graded dual of the Hopf algebra 
$\mathrm{NSym}$ of noncommutative symmetric functions (see e.g. \cite{Malv+Reut95,GKLLRT95,Thibon06}). 
An exploration of the dual of the product $\bu$ would then be a further interesting route 
to pursue.

\end{document}